# Chain Teleportation

Chien-er Lee

Department of Physics, National Cheng Kung University, Taiwan 701

National Center for theoretical Sciences, Taiwan

By means of the idea of measurements on the crossed space-time nonlocal observables, we extend the mechanism for the two-way quantum teleportation to the chain teleportation among N spatially separated spin-1/2 systems. Since in the process only the local interactions are used, the microcausality is automatically satisfied.

In 1993, Bennett, Brassard, Crepeau, Jozsa, Peres, and Wootters (BBCJPW) [1] have proposed a general mechanism to teleport the unknown quantum state at particle 1 to particle 3. To this end the particle 3 is prepared initially in a Bell state with particle 2. Then a Bell state measurement between particles 1 and 2 causes the teleportation (possibly with rotation). In 1997, the first experimental quantum teleportation based on BBCJPW mechanism has been performed by A. Zeilinger et al [2]. In 1998, the first realization of unconditional quantum teleportation was actually teleported by H. J. Kimble et al.[3]. The BBCJPW teleportation may be viewed as one-way teleportation which teleport the state of particle 1 to particle 3, not the reversed. L. Vaidman has proposed another teleportation mechanism (VAA)[4] by extending the method of Aharonov and Albert [5] from nonlocal measurements in space to those in both space and time. This VAA mechanism can achieve a nonlocal swap of states between two spatially separated particles, therefore it is a scheme for two-way teleportations. In this brief repot we extend the two-way teleportation to the chain teleportation of states among N spatially separated spin 1/2 systems. In the following we first briefly review the two-way teleportation, and then describe the mechanism for chain teleportation.

For two spatially separated spin-1/2 systems, in order to achieve the two-way teleportation, the VAA model tries to measure the following "crossed" space-time nonlocal operators

$$[\sigma_{1x}(t_1) - \sigma_{2x}(t_2)] \bmod 4$$

$$[\sigma_{1y}(t_2) - \sigma_{2y}(t_1)] \bmod 4 . \quad (1)$$

where the operator subindices 1 and 2 denote the locations 1 and 2, respectively, and $t_2 > t_1$. In this model only local interactions are used. The Hamiltonian is

$$H = g(t - t_1)[P_1 \sigma_{1x} + P_2 \sigma_{2y}]$$

$$+ g(t - t_2)[P_1' \sigma_{1y} + P_2' \sigma_{2x}] \quad (2)$$

where the normalized $g(t)$ has a compact support around zero; $P_1$ and $P_2$ are conjugate momenta of the pointer variables $Q_1$ and $Q_2$ of devices which locally interact at locations 1 and 2, respectively, at time $t_1$. Similarly, for $P_1'$ and $P_2'$ at time $t_2$. The initial sates of the devices are set to be entangled

$$[Q_1(t_0) - Q'_2(t_0)] \mod 4 = 0, \quad P_1(t_0) + P'_2(t_0) = 0$$

$$[Q'_1(t_0) - Q_2(t_0)] \mod 4 = 0, \quad P'_1(t_0) + P_2(t_0) = 0$$

$$P_i \mod(2\pi\hbar/4) = 0, \quad P'_i \mod(2\pi\hbar/4) = 0 \quad (3)$$

After the interactions are completed, with $T > t_2 > t_1$, we have

$$[\sigma_{1x}(t_1) - \sigma_{2x}(t_2)] \mod 4 = [Q_1(T) - Q'_2(T)] \mod 4$$

$$[\sigma_{1y}(t_2) - \sigma_{2y}(t_1)] \mod 4 = [Q'_1(T) - Q_2(T)] \mod 4 \quad (4)$$

and therefore, local measurements of $Q_i$ and $Q'_i$ whose results are exchanged between locations 1 and 2 yield the values of operators in (1). If the results are zeros, then the two-way teleportation is completed. If they are not zeros, then appropriate local rotations should be performed to achieve it. Since in the process only local interactions are used and only local measurements of devices are performed, the microcausality is automatically satisfied.

Now we try to extend the mechanism for the two-way quantum teleportation to the chain teleportation among N spatially separated spin-1/2 systems. By the chain teleportation we mean to cyclically permute the spin states of N spin-1/2 systems which are located at various different locations. That is if initially the state is at $|\phi_1\rangle_1 |\phi_2\rangle_2 \cdots |\phi_N\rangle_N$, then we want to achieve the final state $|\phi_N\rangle_1 |\phi_1\rangle_2 \cdots |\phi_{N-1}\rangle_N$ by using the method of VAA mechanism. The crossed space-time nonlocal observables to be measured are chosen to be

$$[\sigma_{ix}(t_1) - \sigma_{i+1,x}(t_2)] \mod 4, \quad 1 \leq i < N$$

$$[\sigma_{ky}(t_1) - \sigma_{k+1,y}(t_2)] \mod 4, \quad 2 \leq k < N$$

$$[\sigma_{Nz}(t_1) - \sigma_{1z}(t_2)] \mod 4 \quad (5)$$

where $i(k)$ is the odd(even) number and represents the location. The Hamiltonian is

$$H = g(t - t_1)\left[\sum_{i=1}^{<N} P_i \sigma_{ix} + \sum_{k=2}^{<N} P_k \sigma_{ky} + P_N \sigma_{Nz}\right]$$

$$+ g(t - t_2)\left[\sum_{i=1}^{<N} P'_{i+1} \sigma_{i+1,x} + \sum_{k=2}^{<N} P'_{k+1} \sigma_{k+1,y} + P'_1 \sigma_{1z}\right] \quad (6)$$

where $g$, $P_i$, and $P_i'$ are defined as those in (2) and $t_2 > t_1$. Therefore N devices interact locally with these N systems at $t_1$, and N primed devices interact at $t_2$. The initial sates of the devices are set to be

$$[Q_j(t_0) - Q'_{j+1}(t_0)] \mod 4 = 0, \quad P_j(t_0) + P'_{j+1}(t_0) = 0$$

$$, P_j \mod(2\pi\hbar/4) = 0, \quad P'_j \mod(2\pi\hbar/4) = 0$$

$$j = 1, 2, 3, \cdots, N. \quad (7)$$

where the subindex N+1 is defined to be 1. After the interactions are completed, we obtain

$$[\sigma_{ix}(t_1) - \sigma_{i+1,x}(t_2)] \mod 4 = [Q_i(T) - Q'_{i+1}(T)] \mod 4$$

$$[\sigma_{ky}(t_1) - \sigma_{k+1,y}(t_2)] \mod 4 = [Q'_k(T) - Q_{k+1}(T)] \mod 4$$

$$[\sigma_{Nz}(t_1) - \sigma_{1z}(t_2)] \mod 4 = [Q_N(T) - Q'_1(T)] \mod 4 \quad (8)$$

where the subindex $i$ ($k$) is an odd (even) number which is smaller than N. If the outcomes of local measurements of $Q_j$ and $Q'_j$ for all $j=1,2,3\cdots,N$ give zero results to all expressions in (8), then the chain tele-porttation is completed. If some of them are equal to two, not zeros, then appropriate local rotations should be performed to achieve it. The above chain teleportation may be represented by the following figure

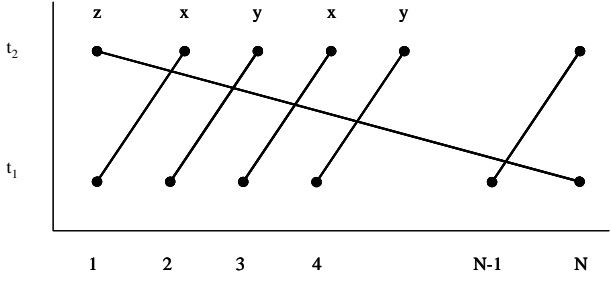

Figure 1. Chain Teleportation : Let *i*(*k*) be the odd(even) number which is smaller than N. Then the chain teleportation is accomplished if $x \equiv \sigma_{ix}(t_1) - \sigma_{i+1,x}(t_2) = 0$, $y \equiv \sigma_{ky}(t_1) - \sigma_{k+1,y}(t_2) = 0$, and $z \equiv \sigma_{Nz}(t_1) - \sigma_{1z}(t_2) = 0$.

In this brief report we have demonstrated how to extend two-way teleportation to chain teleportation by means of crossed space-time nonlocal measurements. We note that the mechanism for chain teleportation is not unique. For example, it is not necessary to use $[\sigma_{Nz}(t_1) - \sigma_{1z}(t_2)] \bmod 4$ in (5). We can just follow the rule to use $[\sigma_{Nx}(t_1) - \sigma_{1x}(t_2)] \bmod 4$ for N is odd, or $[\sigma_{Ny}(t_1) - \sigma_{1y}(t_2)] \bmod 4$ for N is even.